\documentclass{article}
\usepackage{graphicx} 
\usepackage{hyperref} 
\usepackage{orcidlink} 
\usepackage{amsmath}
\usepackage[utf8]{inputenc}
\usepackage{float}

\title{Enhanced Reactivity in Janus Transition Metal Dichalcogenide Quantum Dots: Charge-Density Asymmetry and Hydrodesulfurization Potential}

\author{Jair Othoniel Dominguez Godinez\orcidlink{0009-0009-2352-7087}\thanks{Centro de Investigación Científica y Educación Superior de Ensenada, México, email: jair.dg@cicese.edu.mx} \and 
Raul Eduardo Santoy Flores \footnotemark[5]
\and José Israel Paez Ornelas\orcidlink{0000-0002-3037-5386}\footnotemark[5]  \and 
Rodrigo Ponce Pérez\orcidlink{0000-0002-6726-1569}\footnotemark[5] \and 
Luis Miguel Pellegrin Zazueta\orcidlink{0000-0002-4898-1632}\thanks{Facultad de Ciencias, Universidad Autónoma de Baja California, Mexico} \and
Do Minh Hoat\orcidlink{0000-0003-4835-4505}\thanks{Institute of Theoretical and Applied Research, Duy Tan University, Ha Noi 100000, Vietnam} \thanks{Faculty of Natural Sciences, Duy Tan University, Da Nang 550000, Vietnam} \and
Jonathan Guerrero Sánchez\orcidlink{ 0000-0003-1457-9677}\thanks{Centro de Nanociencias y Nanotecnología, Universidad Nacional Autónoma de México, Mexico, email: guerrero@ens.cnyn.unam.mx}}

\begin{document}

\maketitle
\begin{abstract}
This work investigates the electronic and structural properties of Janus transition metal dichalcogenide quantum dots. We explore various configurations with molybdenum (Mo) and tungsten (W), including oxidized, non-oxidized, and pristine phases. Using computational simulations, we analyze their charge-density asymmetry, thermodynamic stability, and potential applications in hydrodesulfurization reactions.
\end{abstract}
\begin{center}
\end{center}

\section{Introduction}
\subsection{Quantum Dots: Definition and Importance}
Quantum dots (QDs) are nanoscale semiconductor particles that exhibit size-dependent optical and electronic properties due to quantum confinement effects. Their dimensions, typically ranging from 2 to 10 nm, result in discrete energy levels and tunable band gaps, which are not present in their bulk counterparts. These unique quantum effects have positioned QDs as fundamental materials for various technological applications, particularly in optoelectronics, quantum computing, photovoltaics, biomedical imaging, and Catalysis \cite{quantum_dots}. 

The importance of QDs has been recognized globally, leading to the awarding of the 2023 Nobel Prize in Chemistry to Moungi G. Bawendi, Louis E. Brus, and Alexei I. Ekimov for their pioneering work in the discovery and synthesis of quantum dots. Their work demonstrated how QDs could be precisely engineered to exhibit tunable electronic and optical properties, revolutionizing applications in nanotechnology. QDs are currently utilized in high-efficiency light-emitting diodes (LEDs), next-generation displays, quantum computing devices, and even bio-labeling for medical diagnostics \cite{nobel_qds}.

\subsection{Janus Nanoparticles and Their Unique Properties}
Janus nanoparticles are a distinct class of nanomaterials characterized by two chemically or structurally different surfaces, resulting in asymmetric physicochemical properties. This duality allows them to exhibit anisotropic behavior, making them highly versatile for applications in catalysis, drug delivery, and self-assembly processes \cite{janus_nanoparticles}. 

The name "Janus" originates from the Roman god of transitions and duality, symbolizing their two-faced nature. These nanostructures have garnered significant interest in nanoscience due to their ability to achieve selective functionalization, enhance reactivity, and exhibit directional interactions. Janus nanomaterials have demonstrated promising applications in optical imaging, smart materials, and interfacial chemistry \cite{Janus_nanoarchitectures}.

\subsection{Transition Metal Dichalcogenides (TMDs) and Their Janus Phase}
\begin{figure}[ht]
    \centering
\includegraphics[width=1\textwidth]{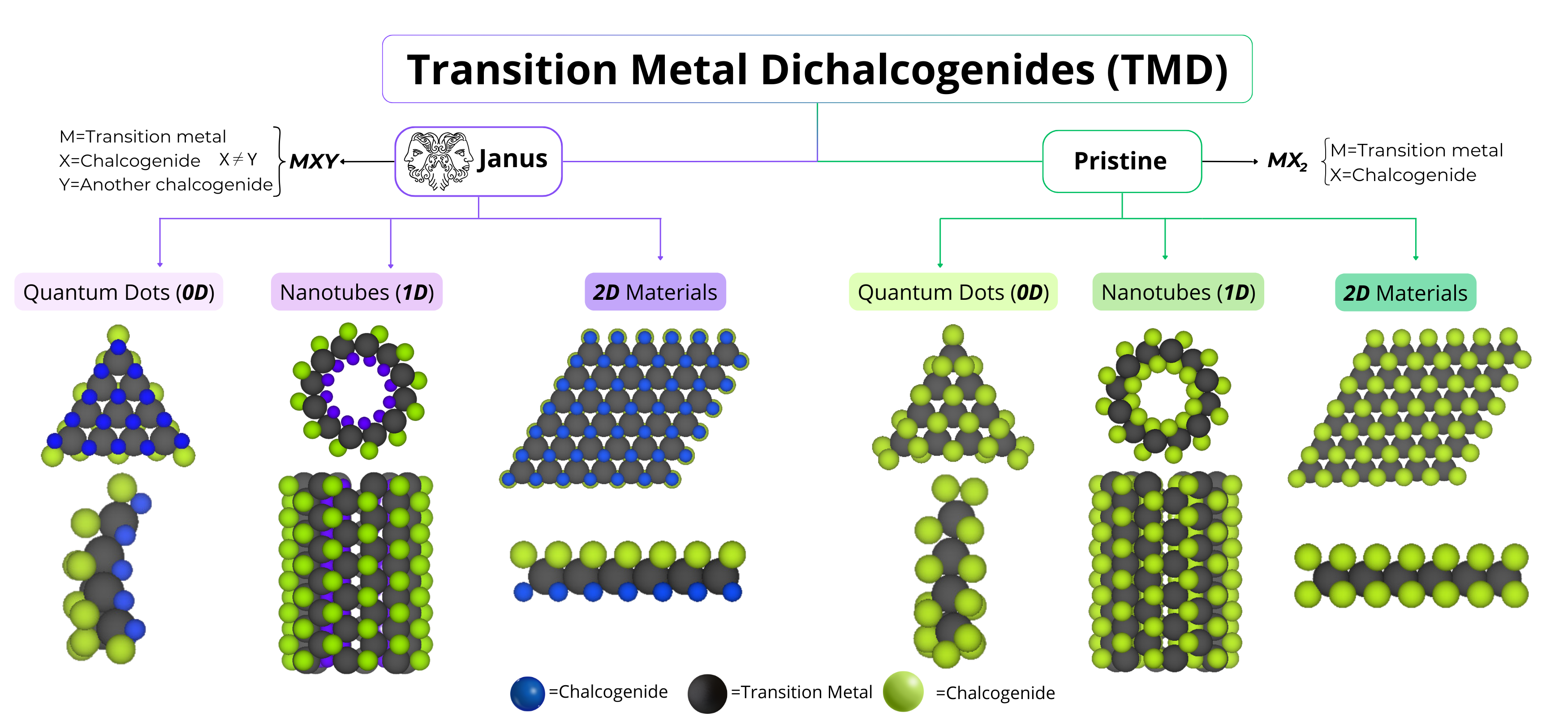}
    \caption{Classification of TMD nanostructures, which are divided into two categories: Janus and pristine. It also provides examples of some possible geometries for these models.}
    \label{fig:tmd_structures}
\end{figure}

Another important family of nanomaterials is the transition metal dichalcogenides (TMDs), which, in their pristine phase, have the chemical formula $MX_2$, where $M$ is a transition metal and $X$ is a chalcogen. These materials can form various nanostructures, including quantum dots with triangular geometry \cite{experimental_mos2_qd}, nanotubes \cite{nanotubos_tmds}, and two-dimensional (2D) monolayers \cite{TMDS}. Figure \ref{fig:tmd_structures} provides an overview of possible TMD nanostructure configurations.

TMDs exhibit a wide range of electronic and catalytic properties, making them highly relevant for applications such as field-effect transistors (FETs) \cite{fet_mos2}, spintronics \cite{spintronics}, photovoltaics \cite{celda.solar.mos2}, and catalysis, particularly in the hydrogen evolution reaction (HER) \cite{tmds_HER}. Their ability to transition between semiconducting, and metallic phases further expands their potential applications.

By selectively replacing one layer of chalcogen atoms in a pristine TMD nanostructure, a Janus TMD is formed, resulting in a chemical formula $MXY$, where $Y$ is a chalcogen different from $X$. Figure \ref{fig:tmd_structures} illustrates different Janus TMD models. This structural modification introduces charge-density asymmetry, thereby breaking inversion symmetry and leading to unique optical and electronic properties. 

For example, monolayers of pristine $MoS_2$ and $MoSe_2$ exhibit direct band gaps of 1.67 eV \cite{moso_mos2_gap} and 1.58 eV \cite{mose2_gap}, respectively. However, their Janus counterparts, $MoSO$ and $MoSeO$, show indirect band gaps of 1.07 eV \cite{moso_mos2_gap} and 0.81 eV \cite{moseo_gap}, making them promising candidates for optoelectronics and solar cell applications \cite{janus_2D}.

\subsection{Triangular Quantum Dots of TMDs and Their Catalytic Potential}
The study of triangular TMD quantum dots (QDs) is particularly relevant due to their potential catalytic applications \cite{HER_nanotriangles}. Among them, $MoS_2$ has been widely investigated both theoretically \cite{mos2_nanotriangles} and experimentally \cite{experimental_mos2_qd}. Research has demonstrated that the morphology and electronic structure of $MoS_2$ nanocrystals strongly depend on their size \cite{size-dependent-mos2}. These triangular QDs can be synthesized via epitaxial growth on Au(111) surfaces \cite{experimental_mos2_qd}.

The edges and vertices of these nanostructures play a crucial role in catalysis. Previous studies have shown that the most stable terminations of $MoS_2$ QDs are sulfur-rich, leading to the formation of one-dimensional metallic edge states \cite{mos2_electronic_edges}. These edge states significantly enhance catalytic activity, particularly in hydrodesulfurization (HDS) reactions, the ability of these QDs to adsorb and react with sulfur-containing molecules has been studied extensively, including the work of Tuxen et al. \cite{size_mos2_dbt}, which demonstrated that $MoS_2$ QDs with fewer than six Mo atoms at the vertices exhibit strong adsorption of dibenzothiophene (DBT).

\subsection{Motivation and Scope of This Work}
Building on these findings, we systematically investigate Janus TMD QDs using first-principles calculations. Specifically, we explore two geometrical models ($\alpha$ and $\beta$) reported in \cite{mosse_quantum_dots} as the most stable structures for Janus QDs. We systematically analyze the electronic and structural properties of $MXY$ quantum dots by varying their size ($n = 4$ to $10$ transition metal atoms along the edge). 

Our study includes:
\begin{itemize}
    \item A comparison betweenoxidized ($MXO$), non-oxidized ($MXY$), and pristine ($MX_2$) QDs.
    \item An analysis of charge-density asymmetry through electrostatic potential surfaces.
    \item The evaluation of surface formation energies to assess thermodynamic stability.
    \item An investigation of the thermodynamic stability through ab initio molecular dynamics of these materials.
\end{itemize}

This article is structured as follows: Section 2 presents the computational details, Section 3 discusses the optimized QDs and their thermodynamic stability, Section 4 characterizes their electronic and structural properties, and finally, Section 5 provides conclusions, potential applications, and future directions.

\section{Computational Details}
\subsection{Methods}

To investigate the properties of Janus TMD quantum dots (QDs), we employed first-principles calculations within the density functional theory (DFT) framework \cite{dft,dft2}. All simulations were performed using the Vienna Ab Initio Simulation Package (VASP) \cite{ab_initio_md,vasp,vasp2}. A simulation cell with dimensions of 24 Å, 23 Å, and 16 Å was used to create vacuum space between periodic images, which was increased proportionally with the QD size to prevent interactions between replicas. 

The electronic states were expanded in plane waves with a cutoff energy of 500 eV for all models. The valence states were treated using the frozen-core approximation within the projected augmented wave (PAW) method \cite{vasp3}. Exchange-correlation interactions were described using the generalized gradient approximation (GGA) with the Perdew–Burke–Ernzerhof (PBE) parameterization \cite{pbe,pbe2}. Structural optimization was carried out until the forces on each atom were below 0.01 eV/Å and the total energy change was less than $1 \times 10^{-4}$ eV. 

The Brillouin zone was sampled using the Gamma scheme. A gamma-centered k-point grid of $1\times1\times1$ was used for the QDs. To compute the chemical potentials of each atomic species, we first considered an $O_2$ molecule in a cubic box of 15 Å with a k-point sampling of $1\times1\times1$. For sulfur, we used the bulk phase with a primitive cell of triclinic geometry and lattice parameters of $a=14.08$ Å, $b=13.56$ Å, and $c=8.49$ Å, with a Brillouin zone sampling of $6\times6\times4$. For selenium, we considered a monoclinic conventional cell with lattice parameters of $a=9.54$ Å, $b=9.54$ Å, and $c=15.47$ Å, and a Brillouin zone sampling of $6\times4\times3$. Finally, for tellurium, we employed a trigonal conventional cell with lattice parameters of $a=b=4.56$ Å and $c=5.90$ Å, with a Brillouin zone sampling of $6\times6\times12$. 

For transition metals ($Mo$ and $W$), a cubic symmetry was assumed, using a k-point grid of $5\times5\times5$ and lattice parameters of $a=3.10$ Å and $a=3.19$ Å, respectively. The bulk phases of pristine TMDs $2H-MX_2$ (where $M=Mo$, $W$ and $X=S$, $Se$, $Te$) were modeled using a k-point sampling of $10\times10\times5$, in agreement with established databases \cite{materials_project}. 

\subsection{Ab Initio Molecular Dynamics (AIMD)}
To assess the thermal stability of the structures, we performed ab initio molecular dynamics (AIMD) simulations using VASP \cite{vasp}. AIMD calculations were conducted at 300 K using the Nosé–Hoover thermostat within an NVT ensemble \cite{aimd1,aimd2}. These simulations allowed us to evaluate the stability of the QDs under thermal fluctuations, providing insights into their structural robustness at ambient conditions.

\subsection{Structural Models}
We considered two geometrical models, $\alpha$ and $\beta$, identified as the most stable structures in \cite{mosse_quantum_dots}. The primary difference between these QD models lies in the number of dimers bonding the transition metal atoms ($Mo$ and $W$) at the edges of the nanotriangles (see Figure 2 for the atomic models). 

To systematically explore size-dependent effects, we generated different sizes for each TMD QD by varying $n$ from 4 to 10, resulting in a total of seven distinct sizes. In total, we modeled 12 Janus TMD QDs ($MXY$) for each geometric configuration. Additionally, the pristine phases ($MX_2$) were also modeled for comparison. 

Altogether, a total of 252 nanostructures were studied, categorized into three groups: oxidized Janus QDs, non-oxidized Janus QDs, and pristine QDs (see Table \ref{tab:quantum_dots_table}).

\subsection{Formation Energy Formalism}
To assess the thermodynamic stability of different sizes and edge terminations of Janus TMD quantum dots (QDs), we employed the formation energy ($FE$) formalism, as defined in Equation \ref{eq:FE}. Our analysis distinguishes between oxidized ($MXO$) and non-oxidized ($MXY$) Janus QDs, using bulk $2H\text{-}MX_2$ structures as a reference to compare the stability of QDs with varying sizes, ranging from $n=4$ to $n=10$ transition metal atoms along the edges. The $FE$ per atom incorporates the chemical potentials of each atomic species in the general formula $M_xX_yY_z$.

\begin{equation}
    FE = \frac{E_{M_xX_yY_z} - x\mu_{M} - y\mu_{X} - z\mu_{Y}}{x + y + z} 
    \label{eq:FE}
\end{equation}

where $E_{M_xX_yY_z}$ represents the total energy of the Janus TMD QD, while $x$, $y$, and $z$ denote the number of each atomic species in the nanotriangle. The chemical potentials of the respective atomic species are denoted as $\mu_{M}$, $\mu_{X}$, and $\mu_{Y}$. 

For oxidized QDs, the formation energy ($FE$) was computed using Equation \ref{eq:FE}. In contrast, for non-oxidized Janus QDs, the surface formation energy (SFE) was computed relative to the bulk phases of TMDs. Thermal equilibrium among the bulk, vacuum, and molecular structures was considered to establish the following correlation among chemical potentials:

\begin{equation}
    \mu_M^{Bulk} + 2\mu_X^{Bulk} + \Delta H_f = \mu_{MX_2} = \mu_M + 2\mu_X
    \label {eq:chemical_potentials}
\end{equation}

where $\Delta H_f$ represents the formation enthalpy of the bulk $MX_2$ reference phase. The chemical potentials of the $X$ species are further constrained by their respective bulk phases:

\begin{equation}
    \mu_X \leq \mu_X^{Bulk}
    \label{eq:chemical_potential_2}
\end{equation}

since 

\begin{equation}
    \mu_{MX_2}^{Bulk} = \mu_{M} + 2\mu_{X}
    \label{eq:chemical_potentials_3}
\end{equation}

Thus, the surface formation energy (SFE) can be rewritten using Equations \ref{eq:FE} and \ref{eq:chemical_potentials_3} as follows:

\begin{equation}
    SFE = \frac{E_{M_xX_yY_z} - x\mu_{MX_2}^{Bulk} + \mu_X (2x - z) - y\mu_Y}{x + y + z}
    \label{eq:SFE}
\end{equation}

Furthermore, for pristine phases, the SFE is calculated using the following expression:

\begin{equation}
    SFE = \frac{E_{M_xX_yY_z} - x\mu_{MX_2}^{Bulk} + \mu_X (2x - z)}{x + y}
    \label{eq:SFE_MX2}
\end{equation}

\begin{table}[h!]
    \centering
    \caption{Simulated QDs, categorized as oxidized, non-oxidized, and pristine.}
    \begin{tabular}{|c|c|c|}
        \hline
        \textbf{Oxidized} & \textbf{Non-oxidized} & \textbf{Pristine} \\ \hline
        $MoSO$ & $MoSSe$ & $MoS_2$ \\ \hline
        $MoSeO$ & $MoSTe$ & $MoSe_2$ \\ \hline
        $MoTeO$ & $MoSeTe$ & $MoTe_2$ \\ \hline
        $WSO$ & $WSSe$ & $WS_2$ \\ \hline
        $WSeO$ & $WSTe$ & $WSe_2$ \\ \hline
        $WTeO$ & $WSeTe$ & $WTe_2$ \\ \hline
    \end{tabular}
    \label{tab:quantum_dots_table}
\end{table}

\begin{figure}[ht]
    \centering 
    \includegraphics[width=1\textwidth]{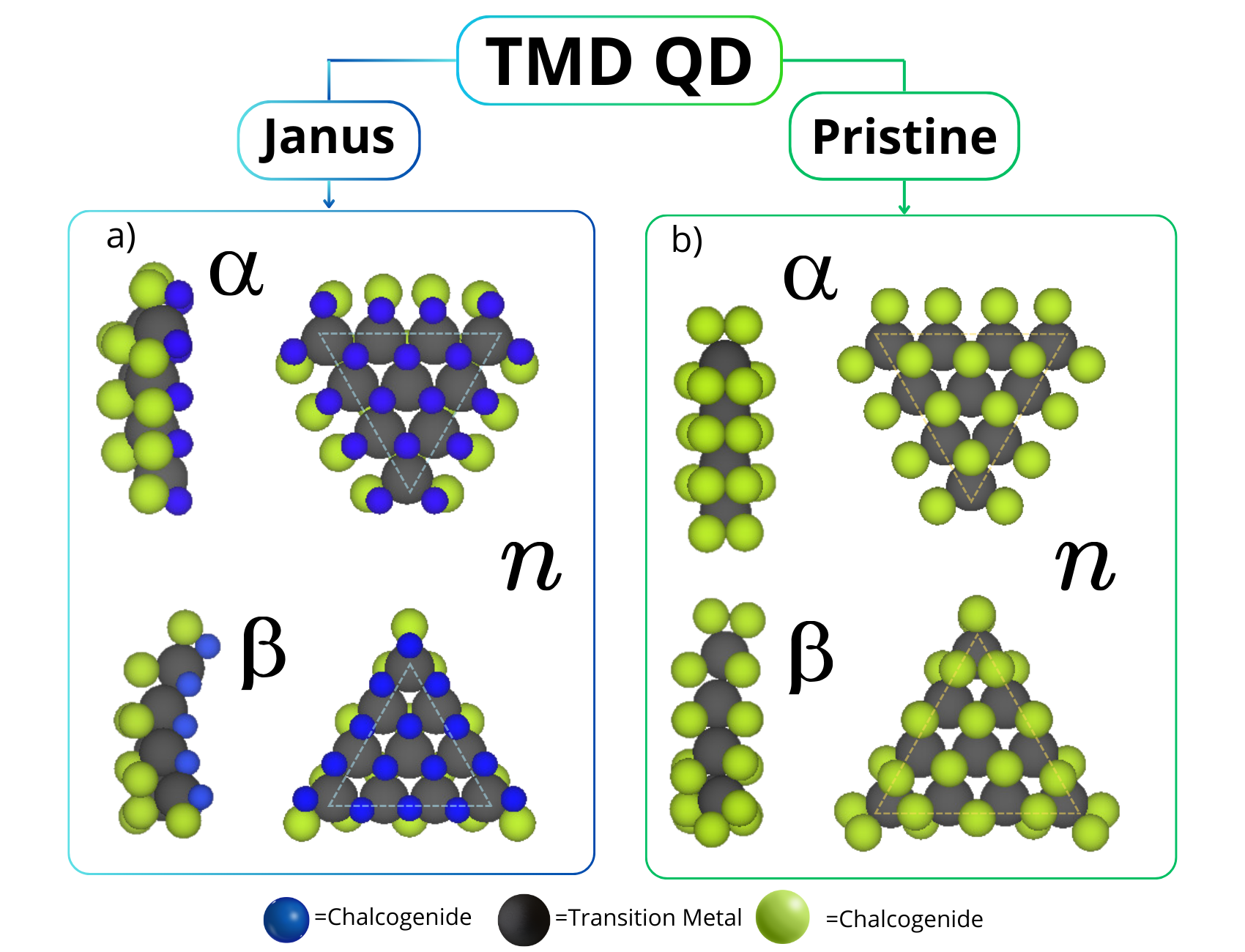}
    \caption{Structural models ($\alpha$ and $\beta$) used to generate the quantum dots. In the $\alpha$ model, the edge transition metal ($M$) atom is bound to a single $X-Y$ dimer, whereas in the $\beta$ model, the edge $M$ atom is bonded to two $X-Y$ dimers. Panel (a) represents the Janus phase, while panel (b) corresponds to the pristine phase. The number of transition metal atoms along the nanotriangle edges is denoted as $n$ and is highlighted with dotted lines.}
    \label{fig:qd-models}
\end{figure}

\section{Results}

\subsection{Optimized Nanostructures}
For each model, structural optimization was performed to determine the minimal energy configuration. After optimization, the relaxed structures were obtained, and it was observed that the oxidized structures exhibit a curvature that increases with the number of transition metal atoms along the edges ($n$). This is a significant result, as it suggests the potential exposure of the basal plane of the Janus TMD QDs, which could increase the number of active sites not only at the edges and vertices but also at the center of the nanotriangle. 

\begin{figure}[p]
    \vspace*{\fill} 
    \centering
    \includegraphics[width=\textwidth]{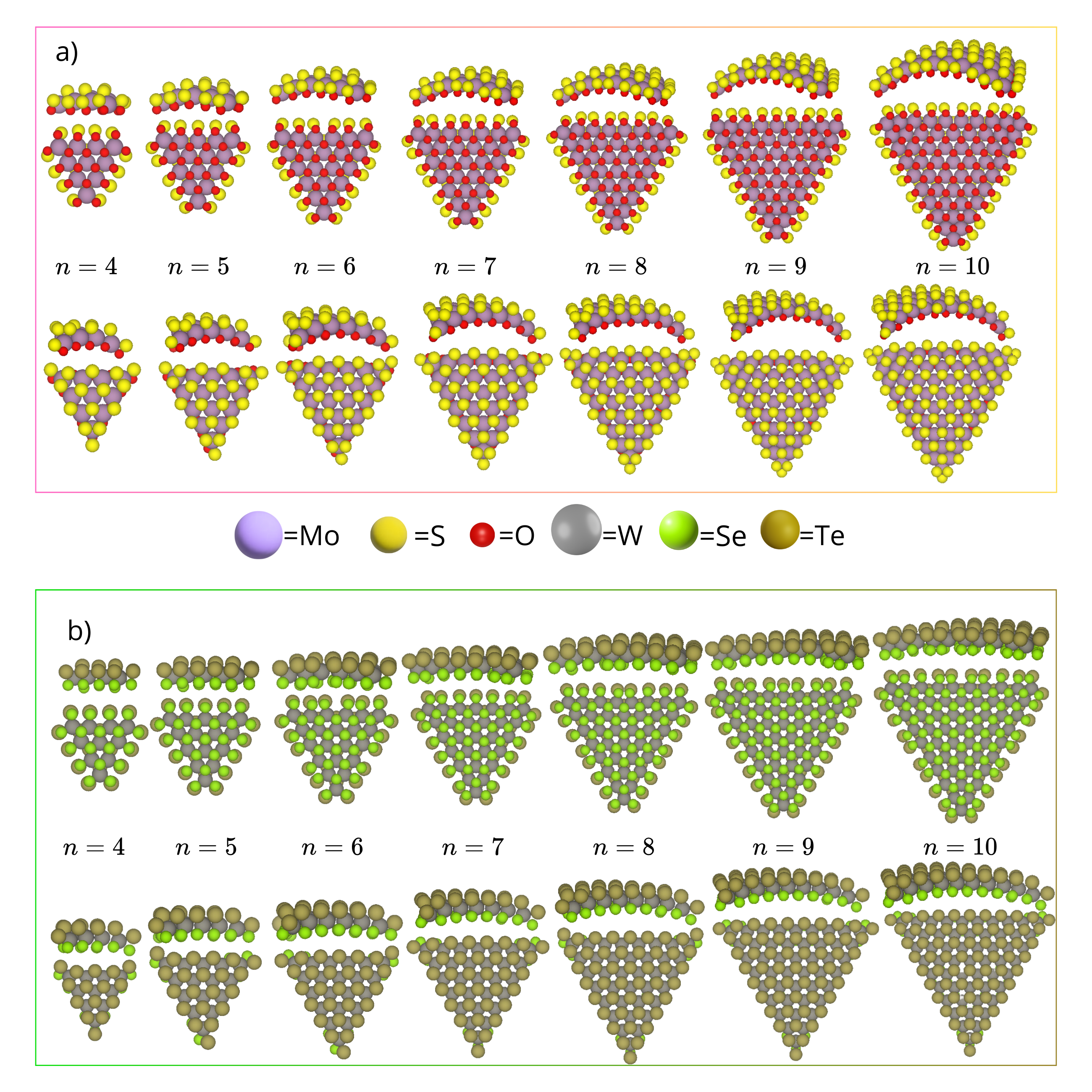}
    \caption{Growth of QDs as $n$ increases, where panel (a) shows $MoSO$ and panel (b) shows $WSeTe$. The oxidized phases exhibit a tendency to become curved with increasing $n$. All these nanostructures were optimized.}
    \label{fig:nanotriangulos}
    \vspace*{\fill} 
\end{figure}

For instance, Figure \ref{fig:nanotriangulos}(a) illustrates the curvature increase for different sizes of nanotriangles in the case of $MoSO$. In contrast, the non-oxidized phase (Figure \ref{fig:nanotriangulos}(b)) and the pristine phases remain nearly flat. A similar trend was observed for the other oxidized TMD QDs, as detailed in the supplementary section.
Curvature Evolution in MoXY and WXY Nanotriangles

As the number of transition metal (TM) atoms at the edges increases, the curvature also increases. This phenomenon suggests that edge strain accumulates progressively, leading to greater out-of-plane distortions. The structural relaxation observed with increasing edge size further confirms the role of atomic interactions in defining curvature trends.
$MoXY$ and $WXY$ nanotriangles exhibit distinct curvature behaviors: $MoXY$ nanotriangles show moderate curvature evolution, indicating that Mo-based structures accommodate strain more flexibly. $WXY$ nanotriangles exhibit stronger curvature effects, possibly due to the influence of heavier W atoms and stronger bonding interactions, which increase resistance to strain relaxation and enhance structural warping. The oxidation state significantly impacts the curvature properties of MoXY and WXY nanotriangles: Pristine $MoXY$/$WXY$: Show controlled curvature due to intrinsic lattice forces maintaining structural stability. Non-oxidized $MoXY$/$WXY$: More flexible, with curvature still increasing as the number of TM atoms at the edges increases. Oxidized $MoXY$/$WXY$: Exhibit higher curvature deviations, likely resulting from surface strain induced by oxygen atoms, which disrupts the equilibrium bonding structure and enhances mechanical deformation. These findings highlight the intricate interplay between composition, oxidation state, and atomic arrangement in determining the curvature properties of transition metal chalcogenide-based quantum dots. Understanding these effects is crucial for tailoring mechanical and electronic properties in nanomaterials for specific applications.
\clearpage
\clearpage
\subsection{Electrostatic potential isosurface analysis}

The electrostatic potential (ESP) isosurfaces provide insight into the charge distribution and potential active sites for chemical interactions in Janus QDs. The analysis was performed for the largest QDs ($n=10$) across both geometrical models ($\alpha$ and $\beta$) and for oxidized, and non-oxidized systems. The $\alpha$ model exhibits electrostatic potential distributions with active sites predominantly located along the edges of the nanotriangle, a behavior consistent across different chalcogen compositions (S, Se, Te) and oxidation states. In contrast, the $\beta$ models exhibit localized charge accumulation in the center of the QDs, likely due to the increased number of transition metal (TM) atoms along the edges, which contribute to a more significant out-of-plane distortion and electron density redistribution. The oxidation state also plays a key role in charge localization. non-oxidized $MoXY/WXY$ QDs exhibit a relatively uniform potential distribution, with mild charge localization at the edges due to intrinsic lattice effects and a moderate charge accumulation, with a trend of increased activity in both edge and central regions, depending on the structural model. Oxidized $MoXO/WXO$ QDs display the most pronounced electrostatic potential variations, particularly in the $\beta$ models, where oxygen atoms introduce significant surface strain, enhancing charge separation and increasing curvature, leading to localized high-potential regions. A comparison between $MoXY$ and $WXY$ systems reveals that $MoXY$ QDs exhibit moderate electrostatic potential gradients, with adaptable charge distribution due to Mo's relatively lower atomic mass and more flexible bonding nature. In contrast, $WXY$ QDs demonstrate stronger charge separation and higher potential contrast, especially in oxidized forms, suggesting that W-based QDs possess stronger bonding interactions, leading to more defined active sites. These findings have significant implications for reactivity and potential applications. The distinct charge localization patterns between $\alpha$ and $\beta$ models suggest that edge or central site activity can be selectively tuned by modifying the QD geometry. Oxidation enhances curvature-induced charge separation, which could be exploited in catalytic applications where active site density is crucial. The higher charge separation observed in $WXY$ systems indicates potential advantages for electronic, hidrodesulfurization and photocatalytic applications compared to their $MoXY$ counterparts. These results highlight the significance of structural design in tuning the electronic and chemical properties of Janus QDs, providing valuable insights for experimental synthesis and functionalization strategies.

\begin{figure}[H]
    \centering
    \includegraphics[width=1.\textwidth]{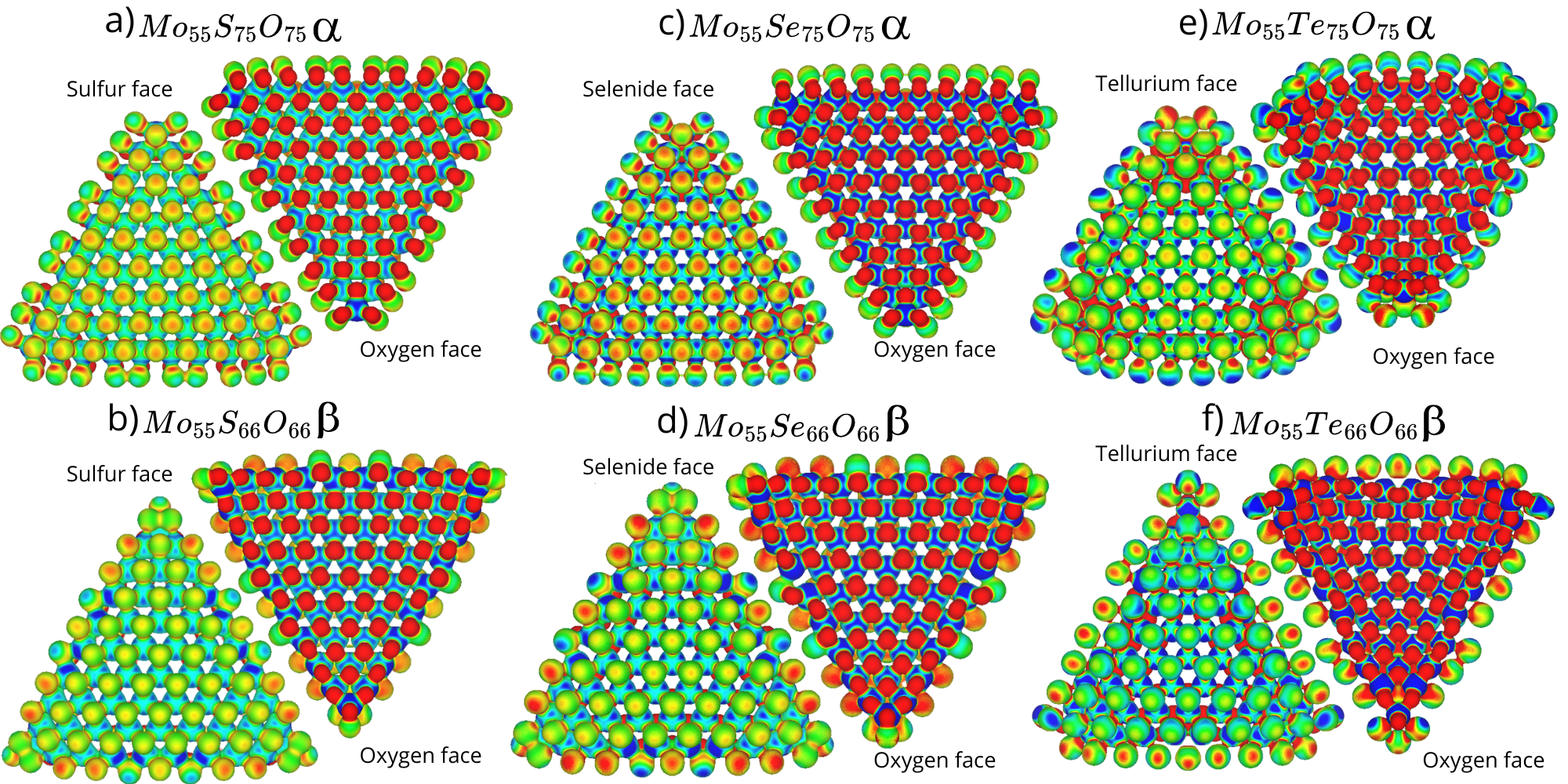}
    \caption{Electroestatic Potential Surfaces of Janus oxides with molibdenum $MoXO$ with a size of $n=10$ }
    \label{fig:EPS_MXO_QDs}
\end{figure}

\begin{figure}[H]
    \centering
    \includegraphics[width=1\textwidth]{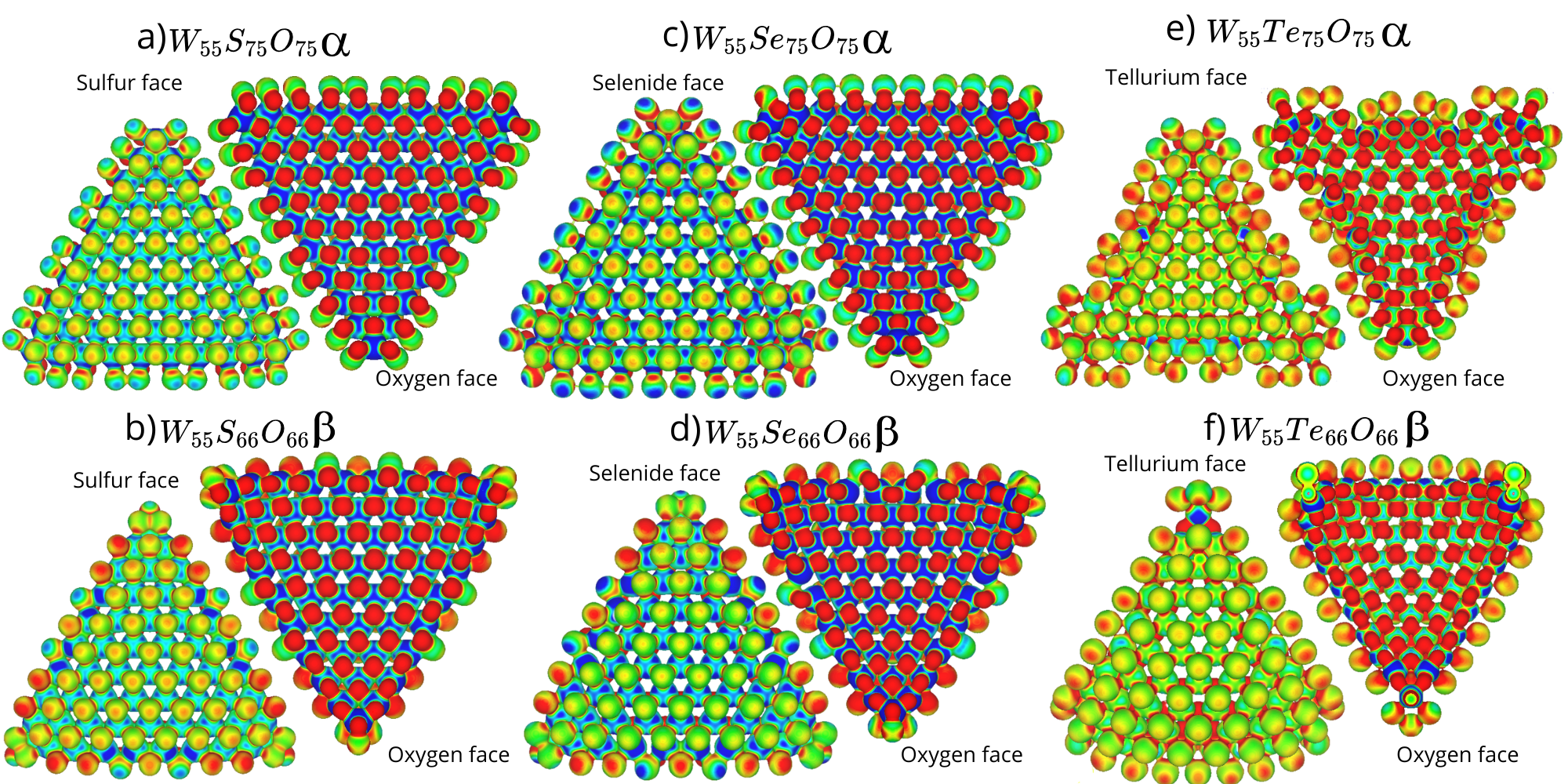}
    \caption{Electroestatic Potential Surfaces of Janus oxides $WXO$ with tungsten a size of $n=10$ }
    \label{fig:EPS_WXO_QDs}
\end{figure}

\begin{figure}[H]
    \centering
    \includegraphics[width=1.\textwidth]{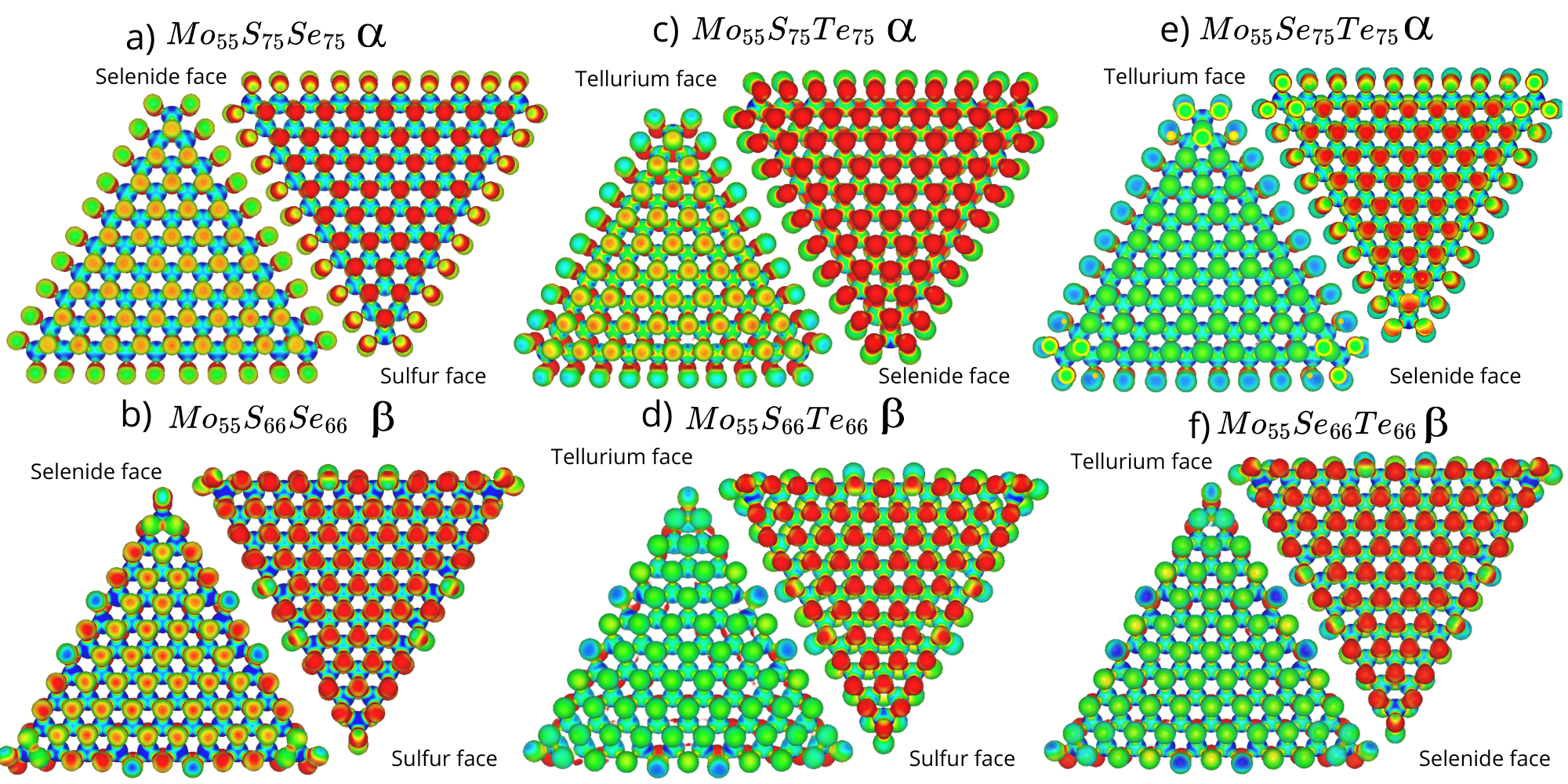}
    \caption{Electroestatic Potential Surfaces of Janus $MXY$ with molybdenum a size of $n=10$ }
    \label{fig:EPS_MXY_QDs}
\end{figure}

\begin{figure}[H]
    \centering
    \includegraphics[width=1.1\textwidth]{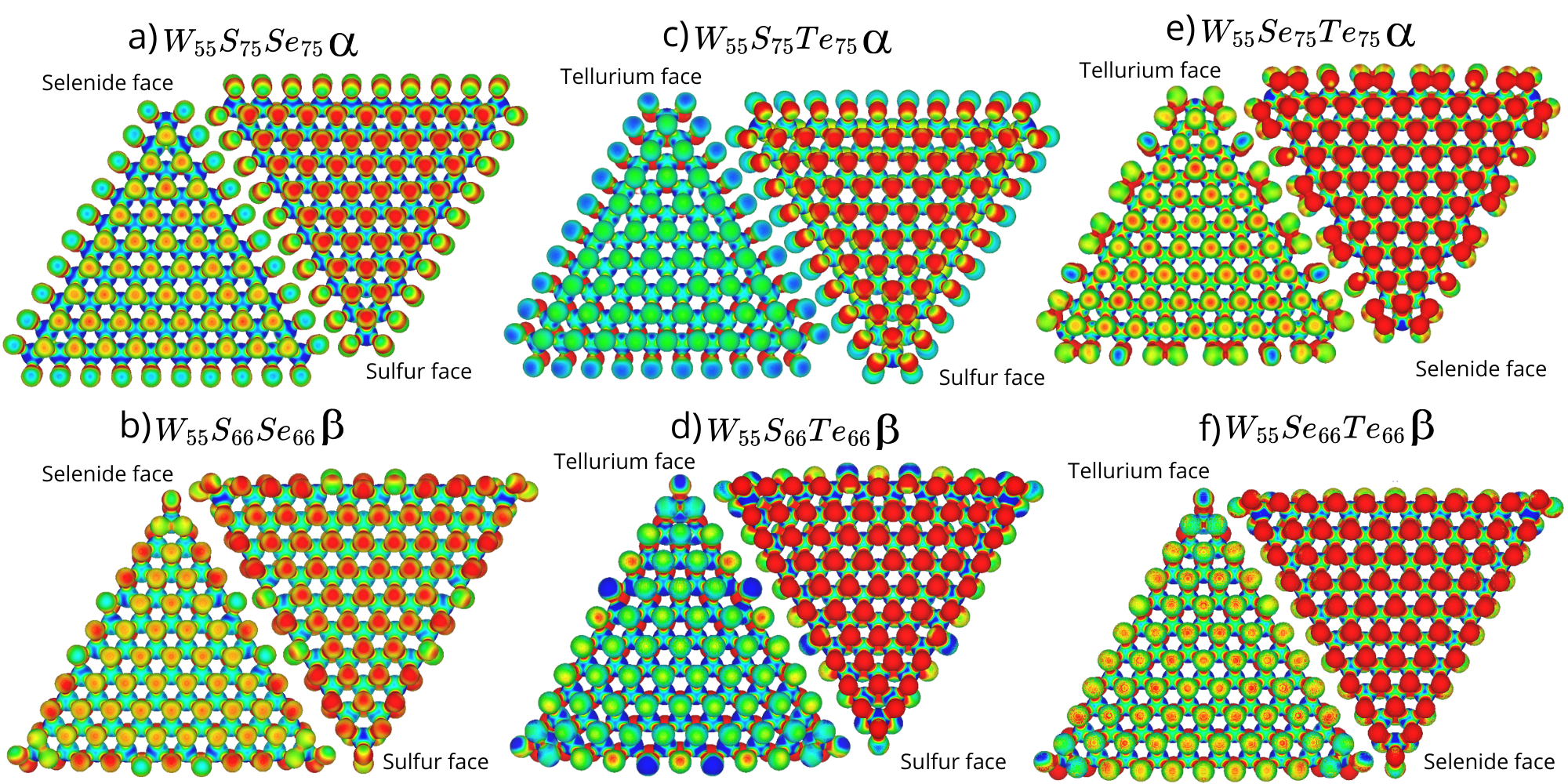}
    \caption{Electroestatic Potential Surfaces of Janus $MXY$ with tungsten a size of $n=10$ }
    \label{fig:EPS_WXY_QDs}
    
\end{figure}

\subsection{Thermodynamic Stability Analysis}
\subsubsection{Analysis of Surface Formation Energies (SFE) for $MXY$ Quantum Dots}

The calculated Surface Formation Energies (SFE) for Mo- and W-based QDs across pristine, oxidized, and non-oxidized phases reveal crucial trends in structural stability. In pristine $MoX_2$ and $WX_2$ QDs, the SFE decreases as $n$ increases, indicating improved stability for larger QDs. Both $MoX_2$ and $WX_2$ exhibit similar trends; however, W-based QDs present slightly lower SFE values, suggesting better thermodynamic stability. Non-oxidized Janus $MXY$ QDs exhibit moderate stability, with SFEs generally higher than those of pristine phases. Additionally, $WXY$ QDs tend to be more stable than $MoXY$, likely due to stronger bonding interactions in W-based systems. Oxidized $MoXO$ and $WXO$ QDs display enhanced stability, with SFEs lower than those of non-oxidized structures, and the stabilization effect is more pronounced in $WXO$ compared to $MoXO$. For all systems (pristine, oxidized, and non-oxidized), increasing $n$ improves structural stability, as larger QDs ($n = 10$) exhibit lower SFEs, confirming that they accommodate strain more effectively. This trend is consistent across Mo- and W-based systems, reinforcing the size-dependent stability observed in previous AIMD simulations. Furthermore, an important result for Janus $MXY$ QDs is that the thermodynamically preferred synthesis pathway depends on the precursor monolayer. In general, it is more favorable to obtain these QDs from the pristine phase with a lower SFE. For example, in the case of $MoSSe$, it is more thermodynamically favorable to synthesize the QD from $MoSe_2$ rather than from $MoS_2$. Similarly, for $WSSe$, the formation from $WSe_2$ is preferred over that from $WS_2$. These results highlight the importance of selecting an appropriate pristine monolayer precursor, as it directly influences the synthesis efficiency and stability of Janus QDs.

\begin{figure}[H]
    \centering
    \includegraphics[width=1.3\textwidth]{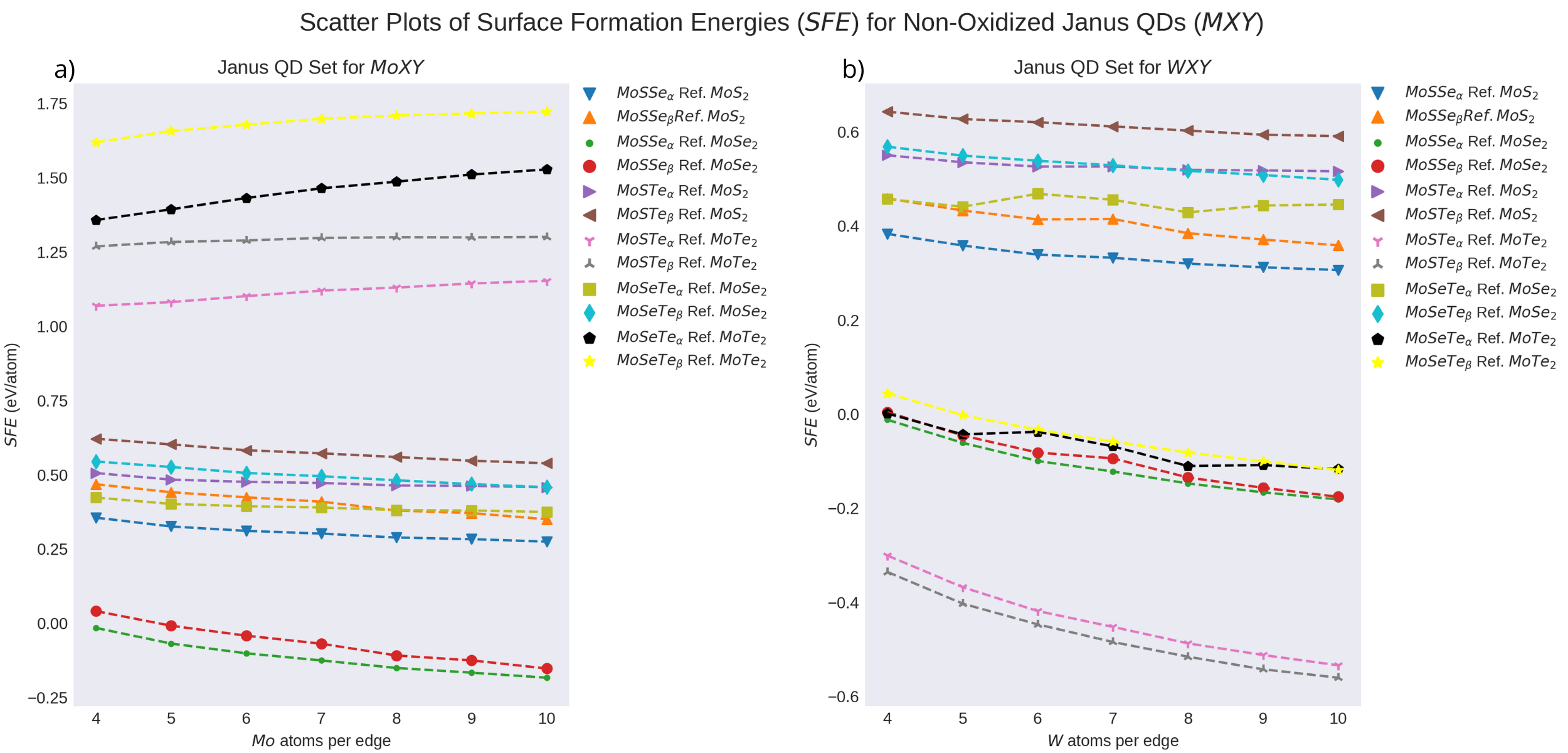}
    \caption{SFE scatter plots for Janus non-oxidized QDs at different sizes: (a) $MoXY$ and (b) $WXY$ families, using bulk references $2H$-$MX_2$ and $2H$-$MY_2$. The surface formation energy (SFE) is calculated using Equation \ref{eq:SFE}.}
    \label{fig:FE_MXY_QDs}
\end{figure}

\begin{figure}[H]
    \centering
    \includegraphics[width=1.3\textwidth]{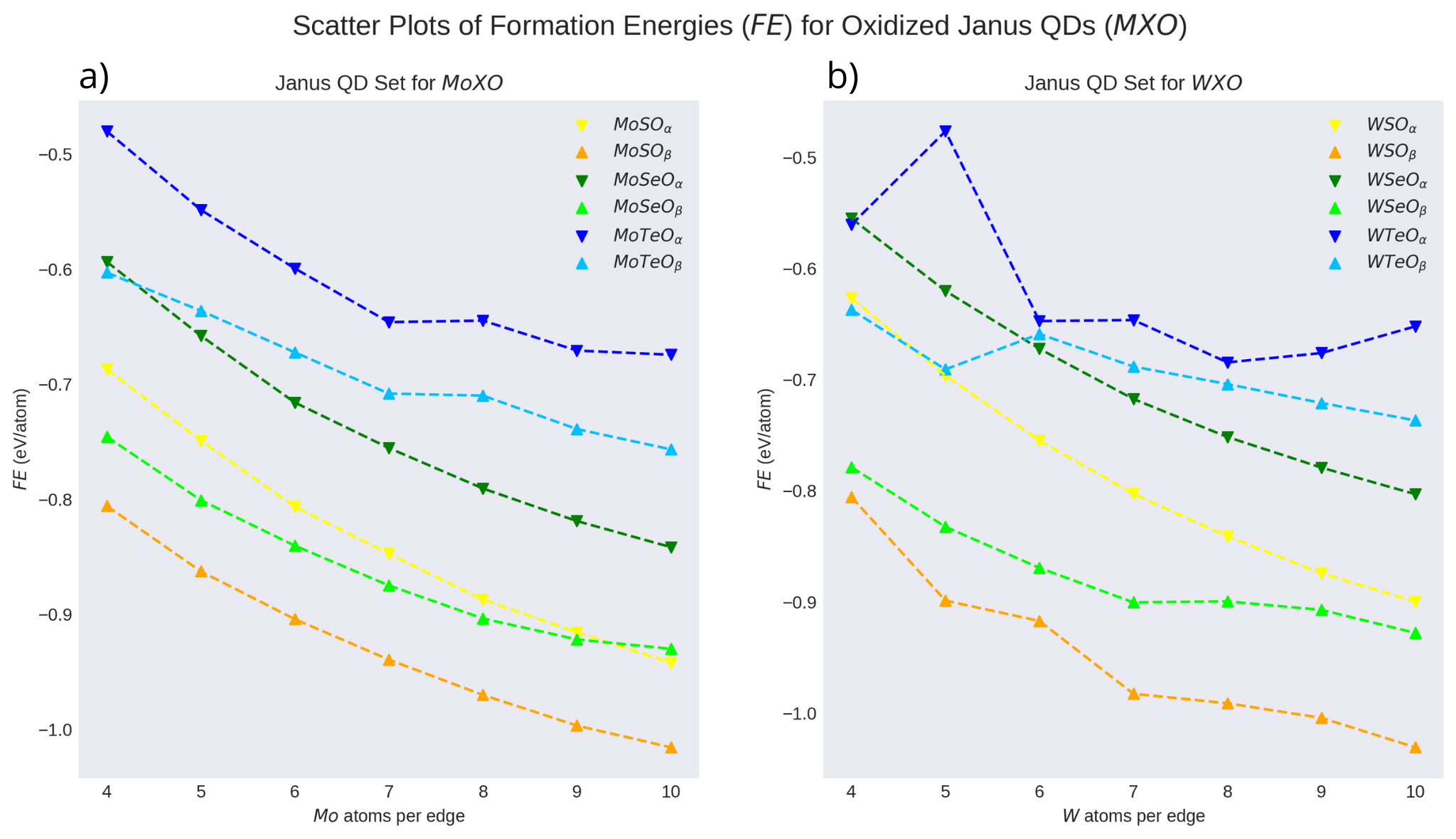}
    \caption{FE scatter plots for Janus oxidized QDs at different sizes: (a) $MoXO$ and (b) $WXO$ families. The formation energy (FE) is computed using Equation \ref{eq:FE}.}
    \label{fig:FE_MOY_QDs}
\end{figure}

\begin{figure}[H]
    \centering
    \includegraphics[width=1.3\textwidth]{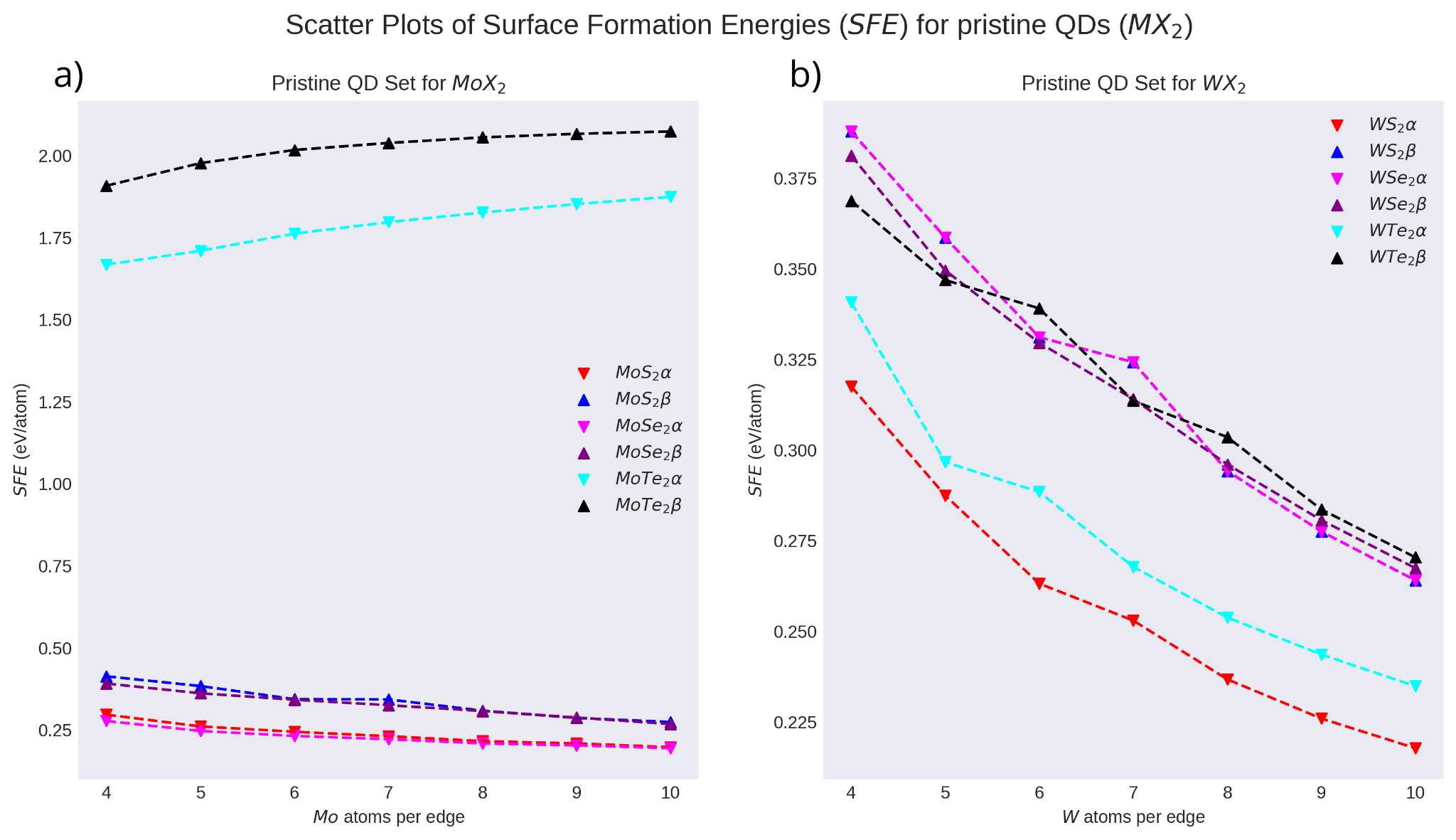}
    \caption{SFE scatter plots for pristine QDs at different sizes: (a) $MoX_2$ and (b) $WX_2$ families. The surface formation energy (SFE) is calculated using Equation \ref{eq:SFE_MX2}.}
    \label{fig:FE_MX2_QDs}
\end{figure}

\subsubsection{Ab initio molecular dynamics}
\subsection{Thermal Stability of $MoSO$ and $WSO$ Quantum Dots}

To investigate the thermal stability of $MoSO$ and $WSO$ QDs, we performed ab initio molecular dynamics (AIMD) simulations using VASP with a simulation time of 5 picoseconds at 300 K. Two geometrical models were considered: $\alpha$ (one monomer at the edges) and $\beta$ (two monomers at the edges), each for $n=4$ and $n=10$. The results reveal a strong size-dependent stability trend, with larger QDs exhibiting greater structural robustness.

For $MoSO$ QDs, the $\alpha$ model exhibits moderate stability at room temperature, with slightly higher energy fluctuations compared to the $\beta$ model. In contrast, the $\beta$ model demonstrates greater structural stability, likely due to increased atomic coordination, leading to reduced fluctuations in both temperature and energy. This suggests that the $\beta$ model is thermodynamically more favorable. A similar trend is observed in $WSO$ QDs, where AIMD simulations indicate that increasing the number of transition metal ($W$) atoms at the edges enhances overall robustness. The smaller $WSO$ QD ($n = 4$) exhibits pronounced fluctuations in temperature and energy but stabilizes toward the end of the simulation, indicating transient instability followed by structural retention at room temperature. In contrast, the larger $WSO$ QD ($n = 10$) demonstrates significantly reduced fluctuations, confirming improved thermodynamic stability.

A clear size-dependent effect is observed for both $MoSO$ and $WSO$ QDs. Smaller QDs ($n=4$) exhibit larger fluctuations in temperature and energy, indicating greater atomic mobility and higher susceptibility to structural distortions. Despite these fluctuations, they remain overall stable. Conversely, larger QDs ($n=10$) exhibit lower energy and temperature variations, confirming superior structural stability. These QDs maintain their triangular shape throughout the simulations, suggesting that size plays a crucial role in enhancing mechanical robustness and thermodynamic stability. The results highlight that larger QDs are promising candidates for experimental synthesis and potential applications, particularly in nanodevices requiring high stability under ambient conditions. The AIMD results for both nanoclusters reveal that the temperature fluctuates around 300 K, indicating a canonical (likely NVT) ensemble, while the total energy stabilizes after initial fluctuations (0.5 ps), suggesting rapid thermal equilibration. The comparison between initial and final atomic configurations shows no significant structural distortions, pointing to good thermal stability, preservation of the $\beta$-phase triangular geometry, and no signs of melting or reconstruction within the 5 ps simulation window. Notably, the larger cluster ($W_{55}S_{66}O_{66}$) exhibits more gradual energy fluctuations, likely due to its higher heat capacity, and appears more structurally robust, while the smaller cluster ($W_{10}S_{15}O_{15}W_{10}$) displays larger energy fluctuation amplitudes, indicating a greater sensitivity to thermal effects.

\begin{figure}[H]
    \centering
    \includegraphics[width=1.3\textwidth]{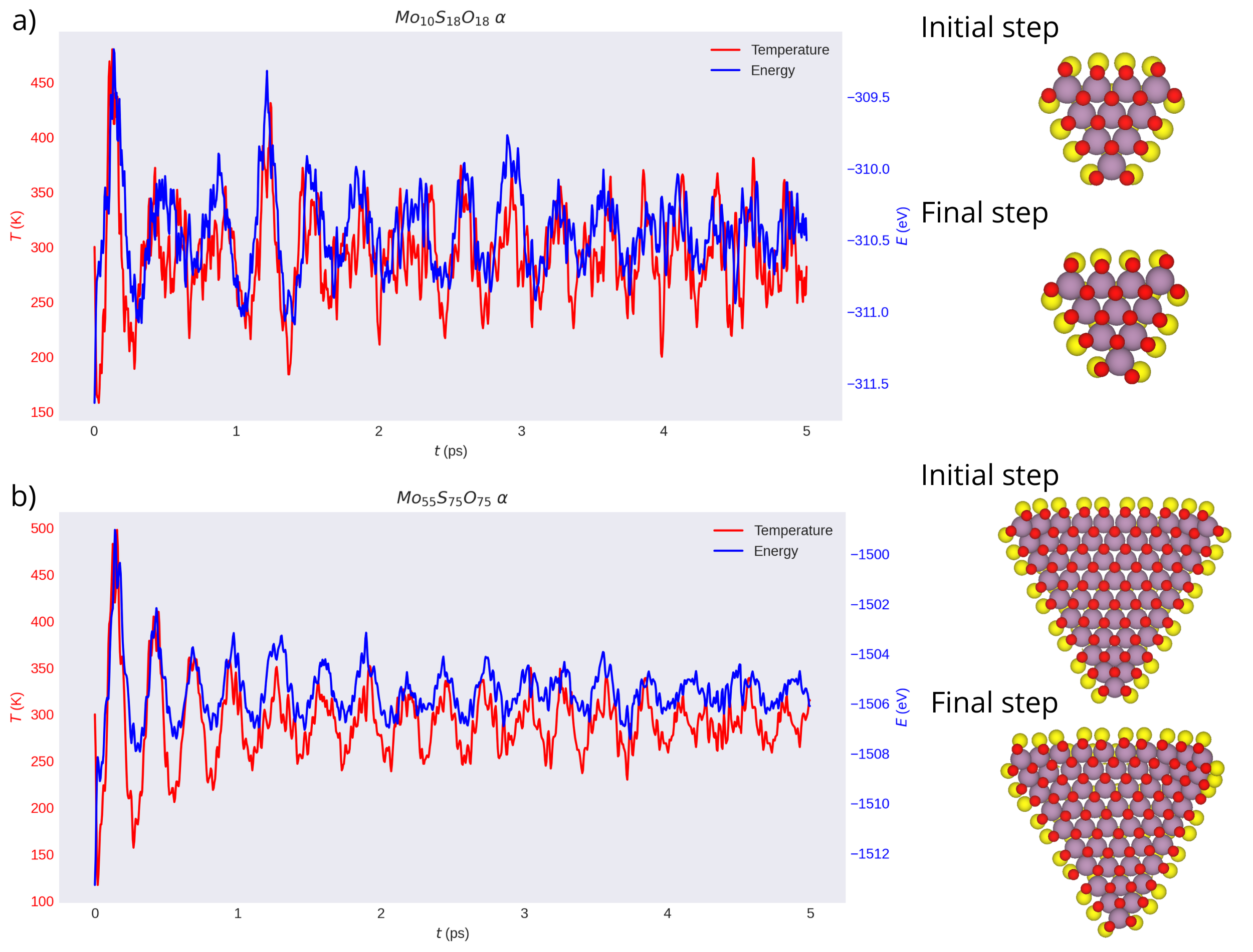}
    \caption{Results of AIMD for the QD $MoSO$ model $\alpha$ for $n=4$ and $n=10$.}
    \label{fig:AIMD_MoSOa}
\end{figure}

\begin{figure}[H]
    \centering
    \includegraphics[width=1.3\textwidth]{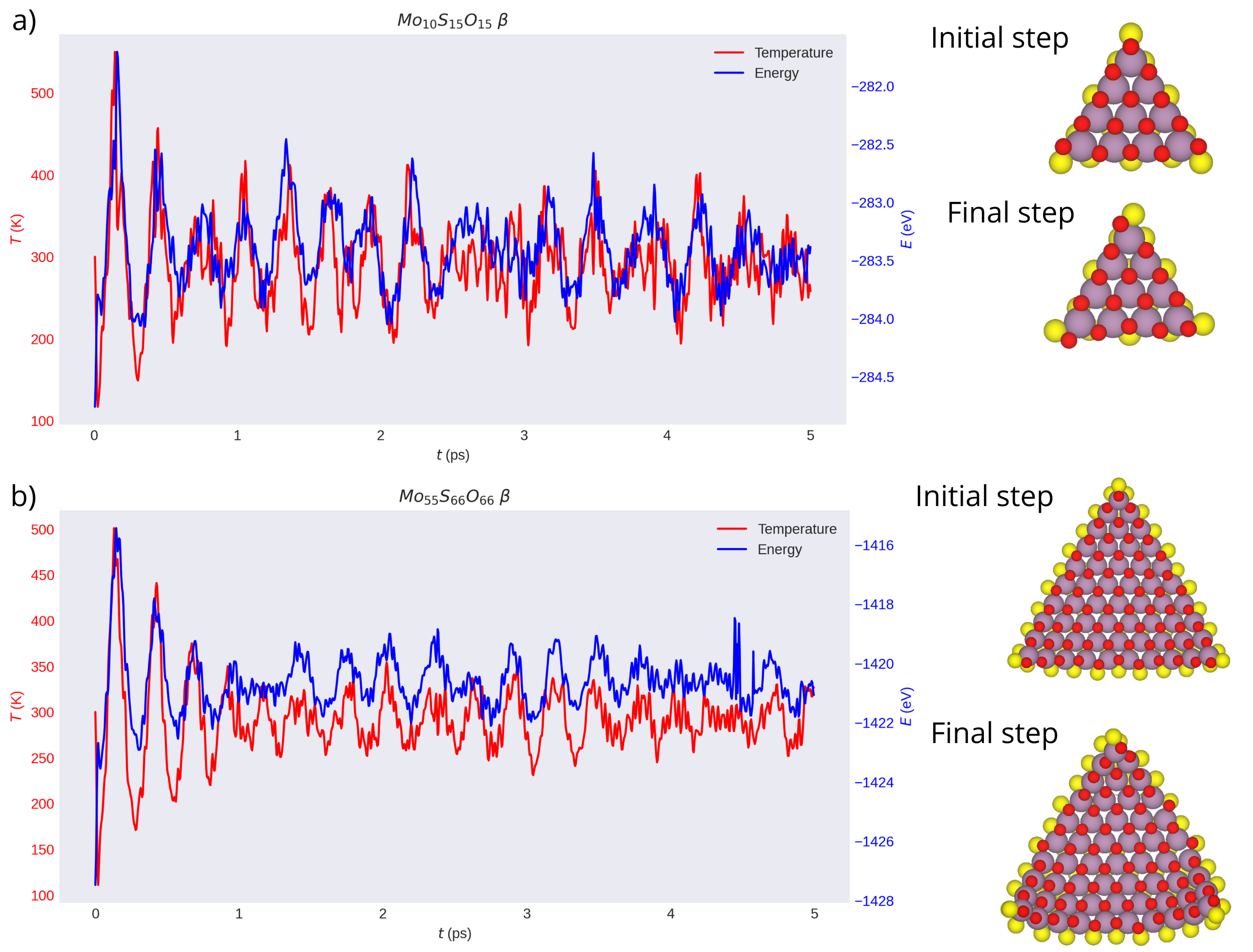}
    \caption{Results of AIMD for the QD $MoSO$ model $\beta$ for $n=4$ and $n=10$.}
    \label{fig:AIMD_MoSOb}
\end{figure}

\begin{figure}[H]
    \centering
    \includegraphics[width=1.3\textwidth]{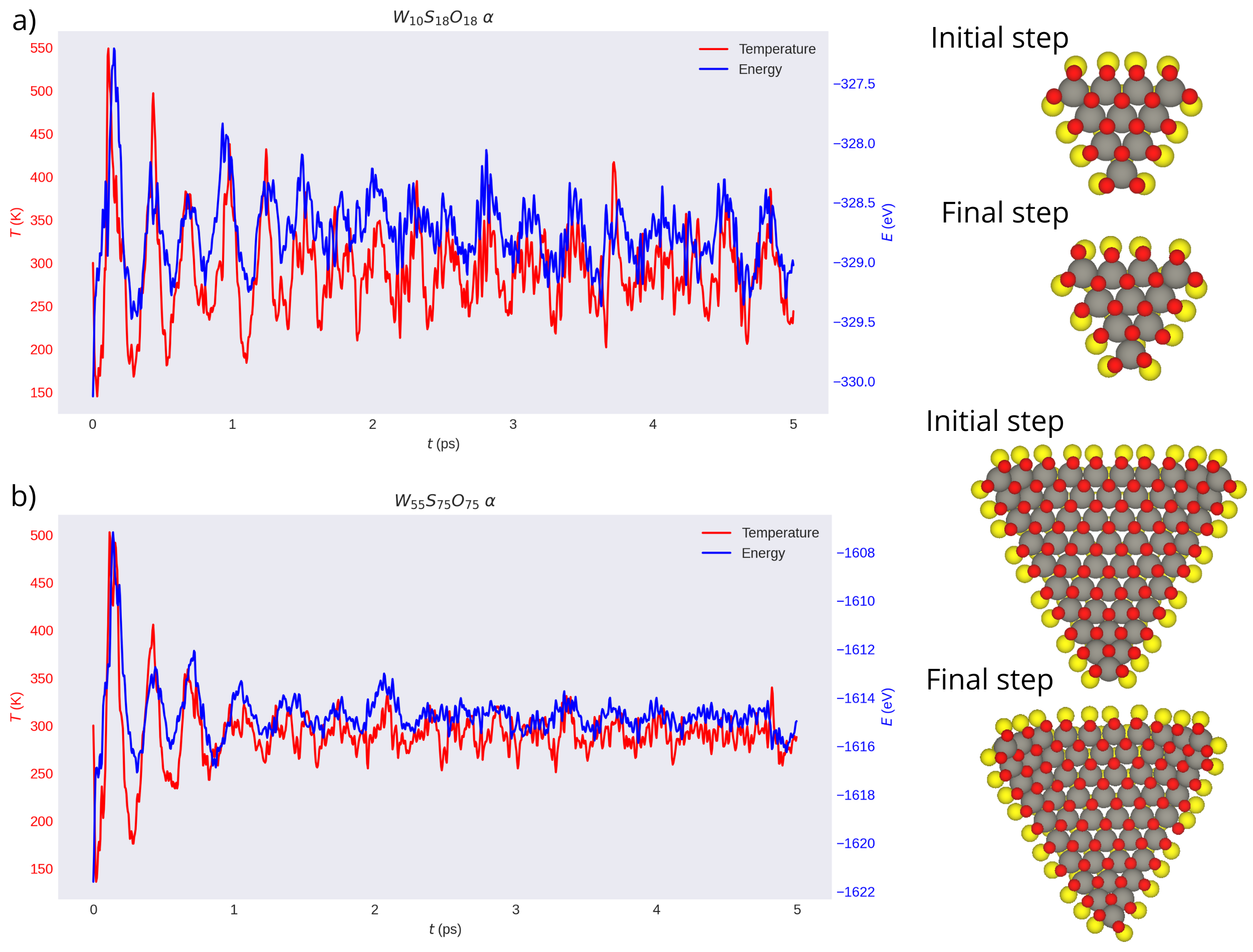}
    \caption{Results of AIMD for the QD $WSO$ model $\alpha$ for $n=4$ and $n=10$.}
    \label{fig:AIMD_WSOa}
\end{figure}

\begin{figure}[H]
    \centering
    \includegraphics[width=1.3\textwidth]{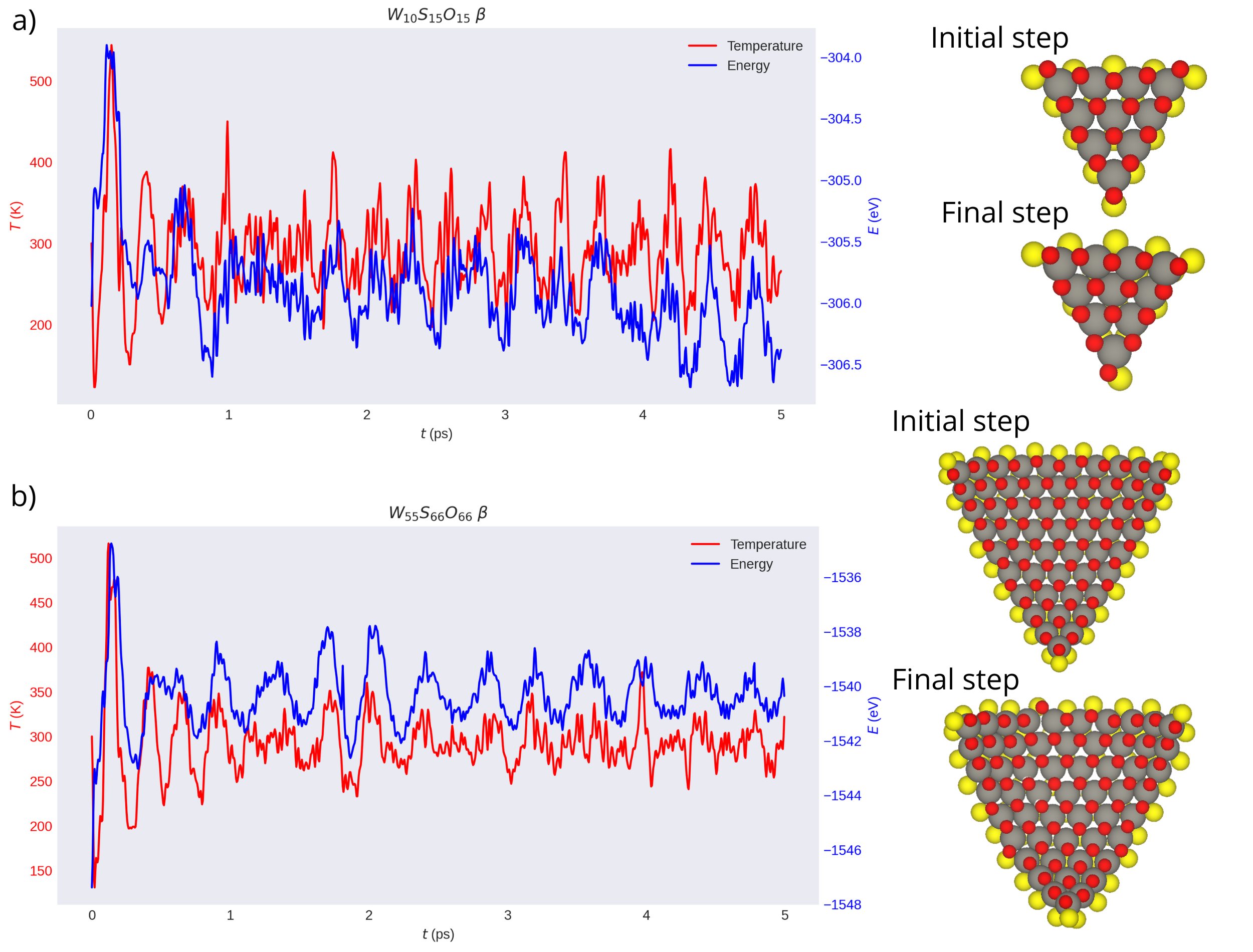}
    \caption{Results of AIMD for the QD $WSO$ model $\beta$ for $n=4$ and $n=10$.}
    \label{fig:AIMD_WSOa}
\end{figure}

\section{Conclusions}

Our study reveals fundamental structural and electronic trends in Janus $MXY$ QDs, highlighting the interplay between size, composition, oxidation state, and geometric configuration. A key structural insight is the evolution of curvature with increasing $n$, where larger QDs exhibit enhanced stability due to improved strain accommodation. $MoXY$ QDs show moderate curvature effects, while $WXY$ QDs display stronger curvature due to the heavier W atoms and their stronger bonding interactions. Oxidation further enhances curvature deviations, increasing surface strain and electronic redistribution. Thermodynamic stability analysis confirms that pristine QDs are the most stable, whereas oxidized QDs demonstrate higher stability than non-oxidized structures. Notably, W-based QDs exhibit lower surface formation energy (SFE) than Mo-based counterparts, indicating superior stability. AIMD simulations confirm that $MoSO$ QDs are thermally stable at room temperature, with the $\beta$ model (two monomers at edges) demonstrating higher robustness than the $\alpha$ model. Larger QDs ($n=10$) maintain greater stability compared to smaller ones ($n=4$), suggesting size-dependent mechanical reinforcement. Electrostatic potential isosurface (EPS) analysis reveals that charge distribution is highly model-dependent; $\alpha$-models exhibit edge-localized charge density, while $\beta$-models display more delocalized charge distributions with active sites emerging at both edges and the center. Oxidation induces charge redistribution, with $MoXY$ QDs showing moderate charge shifts favoring edge activity, whereas $WXY$ QDs exhibit stronger polarization effects due to heavier W atoms. These findings suggest that oxidized $\beta$-models may be promising candidates for oxygen evolution reactions (OER), whereas non-oxidized $\alpha$-models may be better suited for hydrogen evolution reactions (HER) due to their edge-dominated activity. Furthermore, W-based QDs demonstrate potential for nanoelectronic and sensing applications due to their enhanced charge separation. Collectively, these results underscore the importance of geometric model selection, oxidation state, and transition metal identity in tailoring the electronic and catalytic properties of Janus QDs, providing valuable design strategies for future applications in catalysis, nanoelectronics, and sensing technologies.

\newpage
\bibliographystyle{unsrt} 
\bibliography{references.bib} 

\begin{thebibliography}{10}

\bibitem{quantum_dots}
Index.
\newblock In N.~{Thejo Kalyani}, Sanjay~J. Dhoble, Marta Michalska-Domańska,
  B.~Vengadaesvaran, H.~Nagabhushana, and Abdul~Kariem Arof, editors, {\em
  Quantum Dots}, Woodhead Publishing Series in Electronic and Optical
  Materials, pages 581--590. Woodhead Publishing, 2023.

\bibitem{nobel_qds}
K.~David Wegner and Ute Resch‑Genger.
\newblock The 2023 nobel prize in chemistry: Quantum dots.
\newblock {\em Analytical and Bioanalytical Chemistry}, 416:3283,3293, 2024.

\bibitem{janus_nanoparticles}
Xiaoshuang Li, Ligang Chen, Di~Cui, Wei Jiang, Lixia Han, and Na~Niu.
\newblock Preparation and application of janus nanoparticles: Recent
  development and prospects.
\newblock {\em Coordination Chemistry Reviews}, 454:214318, 2022.

\bibitem{Janus_nanoarchitectures}
Ziyang Wua, Li~Li, Ting Liao, Xinqi Chena, Wan Jianga, Wei Luoa, Jianping
  Yanga, and Ziqi Sun.
\newblock Janus nanoarchitectures: From structural design to catalytic
  applications.
\newblock {\em Nano Today}, 22:62,82, 2018.

\bibitem{experimental_mos2_qd}
S.~Helveg, J.~V. Lauritsen, E.~L\ae{}gsgaard, I.~Stensgaard, J.~K. N\o{}rskov,
  B.~S. Clausen, H.~Tops\o{}e, and F.~Besenbacher.
\newblock Atomic-scale structure of single-layer ${\mathrm{mos}}_{2}$
  nanoclusters.
\newblock {\em Phys. Rev. Lett.}, 84:951--954, Jan 2000.

\bibitem{nanotubos_tmds}
S.~Aftab, M.~Iqbal, and Y.~Rim.
\newblock Recent advances in rolling 2d tmds nanosheets into 1d tmds
  nanotubes/nanoscrolls.
\newblock {\em Small}, 19, 2022.

\bibitem{TMDS}
Manish Chhowalla, Hyeon~Suk Shin, Goki Eda, Lain-Jong Li, Kian~Ping Loh, and
  Hua Zhang.
\newblock The chemistry of two-dimensional layered transition metal
  dichalcogenide nanosheets.
\newblock {\em Nature Chemistry}, 2013.

\bibitem{fet_mos2}
Xin Tong, Eric Ashalley, Feng Lin, Handong Li, and Zhiming~M. Wang.
\newblock Advances in mos2-based field effect transistors (fets).
\newblock {\em Nano-Micro Letters}, 2015.

\bibitem{spintronics}
D.~Nguyen, J.~Guerrero, and D.~Hoat.
\newblock Hfxo (x=s and se) janus monolayers as promising two-dimensional
  platforms for optoelectronic and spintronic applications.
\newblock {\em Journal of Material Research}, 2023.

\bibitem{celda.solar.mos2}
Manuel Ramos, John Nogan, Manuela Ortíz-Díaz, José~L Enriquez-Carrejo,
  Claudia~A Rodriguez-González, José Mireles-Jr-Garcia, Roberto~Carlos
  Ambrosio-Lazáro, Carlos Ornelas, Abel Hurtado-Macias, Torben Boll, Delphine
  Chassaing, and Martin Heilmaier.
\newblock Mos2 thin films for photo-voltaic applications.
\newblock {\em IntechOpen}, 2019.

\bibitem{tmds_HER}
Damien Voiry, Jieun Yang, and Manish Chhowalla.
\newblock Recent strategies for improving the catalytic activity of 2d tmd
  nanosheets toward the hydrogen evolution.
\newblock {\em Advanced Materials}, 28:6197,6206, 2016.

\bibitem{moso_mos2_gap}
Tao Wang, Min Su, Hao Jin, Jianwei Li, Langhui Wan, and Yadong Wei.
\newblock Optical, electronic, and contact properties of janus-moso/mos2
  heterojunction.
\newblock {\em The Journal of Physical Chemistry C}, 124:15988,15994, 2020.

\bibitem{mose2_gap}
A.~Kumar and P.~K. Ahluwalia.
\newblock Electronic structure of transition metal dichalcogenides monolayers
  1h-mx2 (m = mo, w; x = s, se, te) from ab-initio theory: new direct band gap
  semiconductors.
\newblock {\em The European Physical Journal B}, 85, 2012.

\bibitem{moseo_gap}
V.~Van On, D.~Khanh Nguyen, J.~Guerrero-Sanchez, and D.~M. Hoat.
\newblock Exploring the electronic band gap of janus moseo and wseo monolayers
  and their heterostructures.
\newblock {\em New J. Chem.}, 45:20776–20786, 2021.

\bibitem{janus_2D}
Hafiza~Sumaira Waheed, Hamid Ullah, M.~Waqas Iqbal, and Young-Han Shin.
\newblock Optoelectronic and photocatalytic properties of mo-based janus
  monolayers for solar cell applications.
\newblock {\em Optik}, 271, 2022.

\bibitem{HER_nanotriangles}
Yurong An, Xiaoli Fan, Hanjie Liu, and Zhifen Luo.
\newblock Improved catalytic performance of monolayer nano-triangles ws2 and
  mos2 on her by 3d metals doping.
\newblock {\em Computational Materials Science}, 159:333--340, 2019.

\bibitem{mos2_nanotriangles}
M.~V. Bollinger, K.~W. Jacobsen, and J.~K. N\o{}rskov.
\newblock Atomic and electronic structure of ${\mathrm{mos}}_{2}$
  nanoparticles.
\newblock {\em Phys. Rev. B}, 67:085410, Feb 2003.

\bibitem{size-dependent-mos2}
J.~Lauritsen, J.~Kibsgaard, S.~Helveg, H.~Topsøe, B.~S. Clausen,
  E.~Lægsgaard, and F.~Besenbacher.
\newblock Size-dependent structure of {MoS\textsubscript{2}} nanocrystals.
\newblock {\em Nature Nanotechnology}, 2:53--58, 2007.

\bibitem{mos2_electronic_edges}
M.~V. Bollinger, J.~V. Lauritsen, K.~W. Jacobsen, J.~K. Nørskov, S.~Helveg,
  and F.~Besenbacher.
\newblock One-dimensional metallic edge states in mos2.
\newblock {\em Physical Review Letters}, 2001.

\bibitem{size_mos2_dbt}
Anders Tuxen, Jakob Kibsgaard, Henrik Gøbel, Erik Lægsgaard, Henrik Topsøe,
  Jeppe~V. Lauritsen, and Flemming Besenbacher.
\newblock Size threshold in the dibenzothiophene adsorption on mos2
  nanoclusters.
\newblock {\em ACS NANO}, 2010.

\bibitem{mosse_quantum_dots}
J.~I. Paez‑Ornelas, R.~Ponce‑Pérez, H.~N. Fernández‑Escamilla, D.~M.
  Hoat, E.~A. Murillo‑Bracamontes, María~G. Moreno‑Armenta, Donald~H.
  Galván, and J.~Guerrero‑Sánchez.
\newblock The effect of shape and size in the stability of triangular janus
  mosse quantum dots.
\newblock {\em Scientific Reports}, 11, 2021.

\bibitem{dft}
W.~Kohn P.~Hohenberg.
\newblock Inhomogeneous electron gas.
\newblock {\em Physical Review}, 1964.

\bibitem{dft2}
W.~Kohn L.~J.~Sham.
\newblock Self-consistent equations including exchange and correlation effects.
\newblock {\em Physical Review}, 1965.

\bibitem{ab_initio_md}
G.~Kresse and J.~Hafner.
\newblock Ab initio molecular dynamics for liquid metals.
\newblock {\em Physical Review B}, 1993.

\bibitem{vasp}
J.~Furthmüller G.~Kresse.
\newblock Efficient iterative schemes for ab initio total-energy calculations
  using a plane-wave basis set.
\newblock {\em Physical Review B}, 1996.

\bibitem{vasp2}
G.~Kresse and J.~Furthmüller.
\newblock Efficiency of ab-initio total energy calculations for metals and
  semiconductors using a plane-wave basis set.
\newblock {\em Computational Materials Science}, 6:15,50, 1996.

\bibitem{vasp3}
S.~Grimme, J.~Antony, S.~Ehrlich, and H.~Krieg.
\newblock From ultrasoft pseudopotentials to the projector augmented-wave
  method.
\newblock {\em Physical Review B}, 59:1758,1775, 1999.

\bibitem{pbe}
John~P. Perdew, Matthias Ernzerhof, and Kieron Burke.
\newblock Rationale for mixing exact exchange with density functional
  approximations.
\newblock {\em The Journal of Chemical Physics}, 1996.

\bibitem{pbe2}
John~P. Perdew, Kieron Burke, and Matthias Ernzerhof.
\newblock Generalized gradient approximation made simple.
\newblock {\em Physical Review Letters}, 1996.

\bibitem{materials_project}
Anubhav Jain, Shyue~Ping Ong, Geoffroy Hautier, Wei Chen, William~Davidson
  Richards, Stephen Dacek, Shreyas Cholia, Dan Gunter, David Skinner, Gerbrand
  Ceder, and Kristin~A. Persson.
\newblock {Commentary: The Materials Project: A materials genome approach to
  accelerating materials innovation}.
\newblock {\em APL Materials}, 1(1):011002, 07 2013.

\bibitem{aimd1}
S.~Nos{\'e}.
\newblock A unified formulation of the constant temperature molecular dynamics
  methods.
\newblock {\em The Journal of Chemical Physics}, 81:511--519, 1984.

\bibitem{aimd2}
W.~G. Hoover.
\newblock Canonical dynamics: Equilibrium phase-space distributions.
\newblock {\em Physical Review A}, 31:1695, 1985.

\end{thebibliography}

\end{document}